\documentclass[conference]{IEEEtran}
\IEEEoverridecommandlockouts
\usepackage{cite}
\usepackage{amsmath,amssymb,amsfonts}
\usepackage{algorithmic}
\usepackage{graphicx}
\usepackage{soul}
\usepackage{textcomp}
\usepackage{xcolor}
\usepackage{xspace}
\usepackage{booktabs}
\usepackage{threeparttable}
\usepackage{multirow}
\def\BibTeX{{\rm B\kern-.05em{\sc i\kern-.025em b}\kern-.08em
    T\kern-.1667em\lower.7ex\hbox{E}\kern-.125emX}}

\definecolor{stelios_colour}{RGB}{200, 238, 200}
\definecolor{light_red}{RGB}{255, 204, 204}

\definecolor{crimson}{rgb}{0.86, 0.08, 0.24}

\newif\ifcomment

\commenttrue

\ifcomment
\newcommand{\stelios}[1]{\sethlcolor{stelios_colour}\hl{[\textbf{Stelios:} #1]}}
\newcommand{\sokratis}[1]{\sethlcolor{orange}\hl{[Sokratis: #1]}}

\else
\newcommand{\stelios}[1]{}
\newcommand{\sokratis}[1]{}

\fi

\makeatletter
\def\footnoterule{\relax%
  \kern-5pt
  \hbox to \columnwidth{\hfill\vrule width 1\columnwidth height 0.4pt\hfill}
  \kern4.6pt}
\makeatother

\newcommand{\tool}{MANE\xspace}

\begin{document}

\title{MANE: A Multi-Path Adaptive Network for Edge Onloading of Deep Neural Networks\\
}

\author{Sokratis~Nikolaidis\IEEEauthorrefmark{2},
        Stylianos~I.~Venieris\IEEEauthorrefmark{3},
        Leonidas~Malachias\IEEEauthorrefmark{2},
        and~Iakovos~S.~Venieris\IEEEauthorrefmark{2}%
\\

\IEEEauthorblockA{\IEEEauthorrefmark{2}National Technical University of Athens, Athens, Greece, \IEEEauthorrefmark{3}Samsung AI Center, Cambridge, UK
}
\IEEEauthorblockA{Email: sokratisnikolaidis@mail.ntua.gr, s.venieris@samsung.com, malachiasnleonidas@gmail.com, venieris@cs.ece.ntua.gr}
\vspace{-0.8cm}
}

\maketitle

\begin{abstract}
    Split computing constitutes a widely used distributed inference approach, where a lightweight head model is onloaded onto the device and a heavier tail model resides on an edge server, leveraging the growing computational capabilities of modern System-on-Chips while alleviating server load. As intelligent indoor environments such as smart offices grow increasingly populated with diverse IoT devices, a single edge server must simultaneously assist multiple devices, each competing for the same shared inference resources. Without a principled mechanism to manage this shared load, the server is quickly overwhelmed, causing latency SLO violations and rendering server-assisted inference ineffective. In this work, we present \tool, a distributed inference framework that equips the server with a multi-path tail architecture, enabling a dynamic accuracy--throughput trade-off at runtime. By introducing a novel multi-path model architecture, a three-stage training scheme featuring a Joint Head Network Distillation loss and a hysteresis-based scheduler with an equitable device-fallback policy, \tool maintains over 80\% SLO satisfaction rate where state-of-the-art onloading methods fail completely, while preserving accuracy 6pp higher than on-device alternatives, across up to 40 concurrent devices.
\end{abstract}


\section{Introduction}
\label{sec:intro}
The proliferation of mobile and Internet-of-Things (IoT) devices has driven a growing demand for deploying sophisticated deep neural networks (DNNs) in smart environments~\cite{laskaridis2024consumer_edge}. Conventionally, to alleviate the heavy computational demands of state-of-the-art DNNs, developers have resorted to cloud or edge offloading, pushing as much computation as possible to a remote server. However, the emergence of modern System-on-Chips (SoCs), equipped with progressively more powerful processors, has expanded the capabilities of embedded devices. 

This shift has led to the new distributed inference paradigm of \textit{onloading}~\cite{dyno2022}. Contrary to offloading, onloading allows server-based DNN applications to deliberately push computation onto the edge devices in order to exploit their growing local compute. In the context of split computing, the server onloads the early stages of computation onto the device, reducing both the server's processing burden and the dimensionality of the intermediate feature maps transmitted over the network. 

While existing split-computing approaches have demonstrated latency and communication benefits, the vast majority assume a static environment where a single device has exclusive access to the server. In realistic smart environments, a single edge server must provide inference assistance to multiple devices concurrently~\cite{nikolaidis2023multitasc}. As the number of assisted devices grows, the aggregate arrival rate of inference requests inevitably surpasses the processing throughput of the server's tail model. In a naive deployment, this contention leads to unboundedly growing queues, causing severe latency service-level objective (SLO) violations. Without a dynamic mechanism to manage the server's load, the system is eventually forced into an indiscriminate fallback to on-device execution, negating the accuracy benefits of server-assisted inference.

To address the limitations of static onloading, we propose \tool,\footnote{Multi-Path Adaptive Network for Edge Onloading.} a novel distributed inference framework designed specifically for multi-tenant edge servers. \tool overcomes rigid throughput constraints by transforming the server-side architecture into a multi-path network, enabling a dynamic trade-off between accuracy and speed. \tool onloads a highly compressed, path-agnostic head to the client devices. During inference, these devices transmit intermediate feature maps to the server, which dynamically routes them through varying depths of the pretrained backbone model based on instantaneous load. To manage this architecture, we introduce a server-side scheduler that continuously monitors request queues, scaling down to faster execution paths during traffic surges, and enforcing a targeted, equitable fallback to on-device execution when maximum capacity is exceeded. The main contributions of the paper are the following:

\begin{itemize}
    \item A novel split-computing architecture that pairs a highly compressed, path-agnostic on-device head with a multi-path server-side tail, introducing a dynamic accuracy–throughput trade-off to edge onloading.
    \item A three-stage training methodology featuring a novel Joint Head Network Distillation loss that mathematically ensures that the onloaded head learns a universally compatible intermediate representation for all server paths.  
    \item A dynamic scheduler that employs hysteresis-based path switching and an equitable device-fallback policy to maximise system-wide accuracy while meeting latency SLOs. 
\end{itemize}

\section{Related Work}
\label{sec:related_work}
\textbf{Split Computing \& Onloading.} The distribution of DNN inference between mobile devices and edge servers has been extensively studied within the scope of split computing~\cite{matsubara2022survey}. After Neurosurgeon~\cite{kang2017neurosurgeon} introduced the partitioning of DNN computation at layer granularity, a large body of work followed, introducing onloading~\cite{dyno2022} and tackling challenges including dynamic split-point selection~\cite{bakhtiarnia2022dynamic, han2022autodidactic}, model partitioning and resource allocation~\cite{optimal2022, clio2020, distredge2022}, and energy-aware inference and feature compression~\cite{autoscale2020, slicer2025, cha2025fast}.


\textbf{Knowledge Distillation for Edge Inference.} Knowledge distillation (KD)~\cite{hinton2015distilling} transfers knowledge from a large teacher model to a compact student by training on the teacher's soft output distributions rather than hard labels alone. Tailored to split computing, Head Network Distillation (HND)~\cite{matsubara2020hnd} distills the on-device head of a bottleneck-injected DNN, minimising both on-device computation and the size of intermediate feature maps transmitted to the server while preserving accuracy. In \tool, we enhance HND in our three-stage training scheme. Subsequent work has further advanced bottleneck-based split computing through improved training strategies, compression benchmarking, and multi-task extensions~\cite{matsubara2022bottlefit, matsubara2023sc2, matsubara2025ladon}. Ladon~\cite{matsubara2025ladon} proposed a single shared encoder compatible with multiple task-specific heads, comprising a parallel to \tool's path-agnostic head design.


\textbf{Multi-Device Scheduling.} The majority of split computing works assume a single device with exclusive access to a server, an assumption that does not hold in real-world, multi-device smart environments. Several works have addressed multi-user inference scheduling at the edge through batching, early exiting, and latency-aware model switching~\cite{shi2023multiuser, she2023multiuser, fluidbatching2023iccad, ecomap2025}. More related to \tool, MultiTASC~\cite{nikolaidis2023multitasc} proposed a multi-tenancy-aware scheduler for cascade architectures that adaptively controls per-device forwarding decisions to maximise throughput while satisfying latency SLOs. \tool focuses on the split-computing setting with a richer accuracy--throughput trade-off enabled through a server-side, multi-path architecture. To the best of our knowledge, this paper is the first to jointly address multi-path split computing, bottleneck distillation, and multi-tenant scheduling in a unified framework.

\section{System Model \& Problem Formulation}
\label{sec:problem_form}
\begin{figure}[t]
    \vspace{0.1cm}
    \centerline{\includegraphics[scale=0.8,trim=0cm 0cm 0cm 0cm]{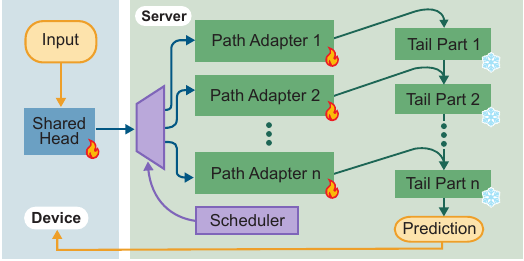}}
    \caption{\small Multi-path adaptive architecture for edge onloading.}
    \label{fig:sys_arch}
    \vspace{-0.4cm}
\end{figure}

In this section, we define the system architecture of the Multi-Device Onloading setting. An edge server seeks to provide DNN inference assistance to a set of $M$ IoT devices $\mathcal{D} = \{d_1, \ldots, d_M\}$, all performing the same task. Rather than processing raw inputs, the server onloads the early stages of computation onto the devices, delegating execution of a lightweight head model to each device and reserving its resources for the more computationally demanding tail.

\noindent\textbf{Single-Device Onloading.} Let $x \in \mathcal{X}$ be a raw input sample on an IoT device. The server onloads a lightweight head model $f_h: \mathcal{X} \rightarrow \mathcal{Z}$ onto the device, which executes it locally to produce a compressed intermediate representation $z = f_h(x) \in \mathcal{Z}$. This representation is transmitted to the server, where the tail model $f_t: \mathcal{Z} \rightarrow [0,1]^K$ completes the inference, producing a probability distribution over $K$ classes, with predicted label $\hat{y} = \arg\max f_t(z)$. Formally:
\begin{equation}
\text{onload}_{f_h, f_t}(x) = f_t(f_h(x))
\label{eq:single_onload}
\end{equation}
From the server's perspective, onloading reduces the volume and dimensionality of incoming data --- the server receives compressed representations $z \in \mathcal{Z}$ rather than raw inputs $x \in \mathcal{X}$ --- alleviating both communication overhead and the server's preprocessing burden. From the device's perspective, the head model is substantially lighter than the full DNN, and raw input data never leaves the device, preserving privacy.

\noindent\textbf{Multi-Device Onloading.} To capture the more realistic setting of multi-device smart environments, we extend the single-device model to $M$ devices, each running the same head model $f_h$ and transmitting compressed representations concurrently. Let $x_m \in \mathcal{X}$ denote the input sample of device $d_m$. The inference result for device $d_m$ is:
\begin{equation}
\text{onload}_{f_h, f_t}(x_m) = f_t(f_h(x_m)), \quad \forall\, d_m \in \mathcal{D}
\label{eq:multi_onload}
\end{equation}
All intermediate representations are placed in a shared request queue $\mathcal{Q}$, from which the server draws batches of size $\mathcal{B}$ and processes them using the shared tail model $f_t$ at a throughput of $\mu_{\mathcal{B}}$ (samples/sec). Each device $d_m$ generates requests at an arrival rate $\lambda_m$, yielding an aggregate arrival rate $\Lambda = \sum_{m=1}^{M} \lambda_m$. As the number of assisted devices grows, $\Lambda$ increases and the server's ability to process requests within a given latency budget degrades. 

\noindent\textbf{Problem Formulation.} The server's objective is to maximise accuracy $\alpha$ across all assisted devices while satisfying a latency SLO $L^*$ for each processed sample. Formally:
%
\begin{equation}
\begin{split}
\max  \alpha \; \text{s.t.} \; & l_{\text{head},m} + l_{\text{transfer},m} + l_{\text{tail}}(\mathcal{B}) \leq L^*, \; \forall\, d_m \in \mathcal{D}
\end{split}
\label{eq:objective}
\end{equation}
where $l_{\text{head},m}$ is the on-device head execution time for device $d_m$, $l_{\text{transfer},m}$ is the network transmission time of the compressed representation from device $d_m$, and $l_{\text{tail}}(\mathcal{B})$ is the server-side tail inference time as a function of batch size $\mathcal{B}$, including queueing delay. While $l_{\text{head},m}$ and $l_{\text{transfer},m}$ remain approximately constant regardless of the number of assisted devices, $l_{\text{tail}}(\mathcal{B})$ grows with $\Lambda$ as requests accumulate in $\mathcal{Q}$ and queueing delay increases.


\section{Proposed System}
\label{sec:architecture}

The multi-device onloading setting demands a server-side architecture capable of dynamically adapting its accuracy--throughput operating point in response to fluctuating demand. To this end, we propose \tool, whose system architecture is illustrated in Fig.~\ref{fig:sys_arch}. Building upon the Head Network Distillation scheme comprising a custom lightweight head with a bottleneck, \tool enhances it by introducing multiple tail adapter paths on the server side, offering multiple options to balance the accuracy--latency trade-off during inference. To ensure stability, high accuracy, low latency and scalable system throughput, \tool uses a three-stage training process and a runtime scheduler that dynamically chooses the execution path and whether to selectively apply on-device execution.



\subsection{Training Scheme}

\begin{figure}[t]
    \vspace{0.1cm}
    \centerline{\includegraphics[scale=0.8]{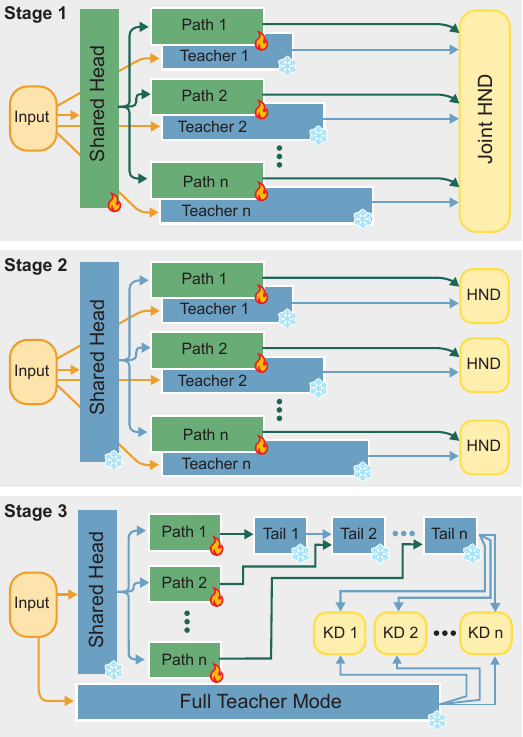}}
    \caption{\small \tool's three-stage training process.}
    \label{fig:training}
    \vspace{-0.4cm}
\end{figure}

Fig. \ref{fig:training} depicts \tool's carefully orchestrated training process for multi-path model splitting, comprising three stages.

\noindent
\textbf{First Stage.} The aim of the first stage is for the student model to learn to produce intermediate feature maps that satisfy the requirements of all paths simultaneously. Rather than training each path to independently mimic its respective teacher, we introduce the Joint Head Network Distillation (JHND) loss. JHND jointly optimises the shared head and all path adapters via a weighted combination of per-path HND losses~\cite{matsubara2020hnd}:
\begin{equation}
\mathcal{L}_{\text{JHND}} = \sum_{i=1}^{N} w_{i} \cdot \| h_{s}^{i} - h_{t}^{i} \|^2
\label{eq:jhnd_loss}
\end{equation}
where $h_{s}^{i}$ and $h_{t}^{i}$ denote the feature maps produced by the student and the respective teacher for path $i$, and $w_{i}$ is a scalar weight controlling each path's contribution to the total loss. This joint optimisation encourages the shared head to learn a path-agnostic intermediate representation. Training utilises early stopping on the validation set. 

\noindent
\textbf{Second Stage.} With the universally compatible representation established, the shared head is now frozen and the pretrained tail remains detached. Each path adapter is trained individually using standard HND loss against its respective teacher head:
\begin{equation}
\mathcal{L}_{\text{HND}}^{i} = \| h_{s}^{i} - h_{t}^{i} \|^2
\label{eq:hnd_loss}
\end{equation}
By isolating each path and freezing the shared representation learned during Stage 1, this second stage allows each adapter to specialise towards its respective teacher signal without interfering with the learned weights of other paths.

\noindent
\textbf{Third Stage.} In the third stage, the shared head remains frozen and the pretrained tail is attached to the pipeline. Each path is then fine-tuned individually using a standard KD loss~\cite{hinton2015distilling}, combining KL divergence between the student and teacher output distributions with a cross-entropy term against the ground-truth labels, balanced by a hyperparameter $b \in [0,1]$ and scaled by temperature $T$. Following this fine-tuning, each path reaches its peak accuracy, matching the performance that would be achieved if trained independently from the outset.

\subsection{Scheduler}

To effectively utilise \tool's multi-path capabilities, we design a server-side scheduler that dynamically adapts the accuracy–throughput trade-off by continuously monitoring system load and via two novel policies: path switching and device fallback. Upon initialisation, the scheduler calculates the largest batch size that can be processed by each path $p$ within the latency SLO, defining this as the path's \textit{capacity}~($C_p$). At run time, the scheduler evaluates the system state by comparing the current request queue length ($QL_{\text{current}}$) against the capacity of the active path ($C_{\text{current}}$).

\noindent
\textbf{Path Switching.} Path switching enables the server to dynamically adapt to fluctuating arrival rates. To prevent rapid oscillation between paths due to instantaneous load spikes, the scheduler employs a hysteresis mechanism by means of a growth counter $GC$ and a drain counter $DC$.

The $GC$ counter tracks sustained queue growth. When the queue length exceeds the current path's capacity and has grown since the last batch was processed, $GC$ is incremented by one; otherwise it is reset to zero:
\begin{equation}
    GC = 
    \begin{cases}
        GC+1 & \text{if}  \quad QL_{\text{current}} > C_{\text{current}} \quad \\
        & \text{and} \quad QL_{\text{current}}>QL_{\text{last}}\\
      ~ ~  0 &  \text{otherwise}
    \end{cases}
    \label{eq:path_switching_faster}
\end{equation}
where $QL_{\text{current}}$ is queue length at scheduling time, $C_{\text{current}}$ the active path's capacity, and $QL_{\text{last}}$ queue length at the previous batch's time of processing. When $GC$$\ge$$GL$, where $GL$ is the growth limit, the scheduler switches to the next fastest path.

Conversely, the drain counter $DC$ tracks sustained queue draining. When the queue length falls below the capacity of the next more accurate path ($C_{\text{accurate}}$) and has decreased since the last batch, $DC$ is incremented; otherwise it is reset to zero:
\begin{equation}
    \it{DC} = 
    \begin{cases}
        DC+1 & \text{if}  \quad QL_{\text{current}} < C_{\text{accurate}} \quad \\
        & \text{and} \quad QL_{\text{current}} < QL_{\text{last}}\\
      ~ ~  0 &  \text{otherwise}
    \end{cases}
    \label{eq:path_switching_accurate}
\end{equation}
where $C_{\text{accurate}}$ is the capacity of the next more accurate path. When $DC \geq DL$, where $DL$ is the drain limit, the scheduler switches to the more accurate path to maximise accuracy.

\noindent


\noindent
\textbf{Device Fallback.} Device fallback is triggered when server demand surpasses the processing capacity of even the fastest path, using the same growth counter logic as path switching with the additional condition that the current path must already be the fastest. When $GC$$\ge$$GL$, a fraction $\phi$ of devices — prioritised by sample submission volume to ensure fair resource allocation — is instructed to process their workloads locally using a lightweight fallback model, directly reducing the total arrival rate $\Lambda$ and restoring the server's ability to meet the latency SLO. Fallback devices are periodically rotated to maintain equitable quality of service under prolonged load, and are reintroduced to server-side onloading one at a time once the queue falls below the fastest path's capacity and exhibits sustained draining, avoiding sudden load oscillations. Unlike path switching, fallback enforcement is subject to a minimum cooldown period $T_{\text{cooldown}}$ between successive engagements.



\section{Evaluation}
\label{sec:evaluation}
\subsection{Experimental Setup}

\begin{table}[t]
    \caption{\small Compression Analysis}
    \centering
    \resizebox{0.475\textwidth}{!}{
    \setlength{\tabcolsep}{2pt}
    \begin{threeparttable}
        \begin{tabular}{llcc}
            \toprule
            \textbf{Model} & \textbf{Path} & \textbf{Param. Reduction} & \textbf{Feature Map Reduction} \\
            \midrule
            \multirow{3}{*}{ResNet152}
                & Accurate  & $\phantom{1}213\times$   & $42.7\times$ \\
                & Balanced  & $1550\times$  & $21.3\times$ \\
                & Fast      & $1922\times$  & $21.3\times$ \\
            \midrule
            \multirow{3}{*}{ConvNeXt Small}
                & Accurate  & $\phantom{1}102\times$   & $16\times$  \\
                & Balanced  & $1329\times$  & $\phantom{1}8\times$   \\
                & Fast      & $1930\times$  & $\phantom{1}8\times$   \\
            \midrule
            \multicolumn{2}{l}{vs. Raw Input (all paths)} & --- & $16\times$ \\
            \bottomrule
        \end{tabular}
    \end{threeparttable}
    }
    \label{tab:compression}
\vspace{-0.4cm}
\end{table}

\tool's multi-path model architecture was implemented using PyTorch 2.11.0. 
We target an edge server equipped with a 10GB NVIDIA RTX 3080 GPU and an AMD Ryzen 5 3600 3.6GHz CPU, for both training and deployment. For the client devices, we target Samsung S20 FE, a dated flagship smartphone to simulate a median-capability user device, with device-side models deployed using PyTorch Mobile targeting the device's NPU. For device--server communication, we employ the AMQP protocol, following standard practice for IoT device messaging.
We target 200-class image classification using the Tiny ImageNet dataset, comprising 100k samples with an 80\%--10\%--10\% training--validation--test split. All deployment results are reported on the held-out test set.

\noindent\textbf{Hyperparameters.}
All stages use the AdamW optimiser with early stopping on the validation set with a patience of 3. The learning rate is set to $10^{-3}$ for Stages 1 and 2, and reduced to $10^{-4}$ for Stage 3. Stage-specific hyperparameters are set as follows: $w_i = \frac{1}{N}$ for Stage 1's JHND, $b = 0.9$ and $T = 4$ for Stage 3's KD. The scheduler growth $GL$ and drain limits $DL$ are set to 3 and 5, respectively, and $\phi$ to 0.2.

\noindent\textbf{Evaluation Protocol.}
We evaluate \tool on ResNet152~\cite{he2016resnet} and ConvNeXt Small~\cite{liu2022convnext} adaptations, targeting latency budgets ranging from 100ms to 200ms with a varying number of devices. A \textit{three}-path structure was implemented for each backbone model. The initial split point was chosen to preserve full accuracy, while the two additional tail injection points were selected based on the accuracy--throughput curve of candidate injection points to realise an \textit{accurate--balanced--fast} path scheme. For ResNet152, the \textit{balanced} and \textit{fast} injection points were placed 14 and 18 blocks after the initial split point, respectively, while for ConvNeXt Small at 12 and 18 blocks. 

Each device is assigned the same subset of 1000 images, randomly sampled from the test set using a fixed seed. Each experiment is repeated across three different seeds, reporting average, minimum, and maximum values. The sample order is shuffled before each run, and inter-arrival delays are sampled from a Poisson distribution with a mean of 20~ms to emulate realistic device usage patterns. Device inference latency measurements use a batch size of 1, averaged across 500 runs. The evaluation metrics are system throughput, average accuracy across devices, and SLO satisfaction rate, which is the percentage of samples processed within a latency target.
\begin{table}[t]
    \caption{\small Path Specifications}
    \centering
    \setlength{\tabcolsep}{3pt}
    \begin{tabular}{llcc}
        \toprule
        \textbf{Model} & \textbf{Path} & \textbf{Accuracy} & \textbf{Accel.} \\
        \midrule
        \multirow{3}{*}{ResNet152}
            & Accurate  & 83.98\% & $1.23\times$ \\
            & Balanced  & 80.79\% & $1.85\times$ \\
            & Fast      & 78.03\% & $2.11\times$ \\
        \midrule
        \multirow{3}{*}{ConvNeXt Small}
            & Accurate  & 89.39\% & $1.35\times$ \\
            & Balanced  & 85.43\% & $2.12\times$ \\
            & Fast      & 80.39\% & $2.43\times$ \\
        \bottomrule
    \end{tabular}
    \label{tab:paths}
\vspace{-0.6cm}
\end{table}

\noindent\textbf{Baselines.}
We compare \tool against two baselines: \textit{i)}~server-only execution, where samples are processed by the original pretrained model without any path switching or fallback, and \textit{ii)}~on-device execution, where all inference is performed locally using MobileNetV2~\cite{sandler2018mobilenetv2}. We also use two ablated versions of \tool to showcase the importance of each component: \textit{1)}~\tool-NPS, where we disable both path switching and fallback from the scheduler, allowing only the \textit{accurate} path to be used, and \textit{2)}~\tool-NF, where we enable path switching and remove fallback.

\subsection{Model Evaluation}

\noindent\textbf{Compression Analysis.}
Table~\ref{tab:compression} shows the compression achieved by \tool's shared head relative to the original backbone models. The head model comprises approximately 12.6k parameters, representing a reduction of $213\times$ and $102\times$ over the \textit{accurate}-path split point of ResNet152 and ConvNeXt Small, respectively. The intermediate feature maps produced by the head have a shape of $[12 \times 28 \times 28]$, yielding a transmission cost reduction of $\approx 16\times$ relative to the raw input and up to $\approx 42.7\times$ relative to the naive split point of ResNet152's fast path. As a result, \tool significantly reduces both the on-device compute cost and the communication overhead, independently of which server-side path is active.

\noindent\textbf{Per-Path Performance.}
Table~\ref{tab:paths} reports the accuracy and acceleration of each path over the pretrained backbone, using a batch size of 128. The three paths offer a consistent accuracy--throughput trade-off across both model families. The three-stage training scheme manages to preserve accuracy across all paths. Notably, the \textit{accurate} path matches or slightly exceeds the pretrained backbone's accuracy, demonstrating that the JHND-based training in Stage 1 successfully guides the shared head towards a representation that serves all paths without compromising \textit{accurate}. \textit{Balanced} and \textit{fast} modestly trade accuracy for substantial speedup, with \textit{fast} delivering up to $2.11\times$ for ResNet152 and $2.43\times$ for ConvNeXt Small.

\noindent\textbf{End-to-end Model Performance.}
Table~\ref{tab:models} compares \tool against the original pretrained models in terms of accuracy and inference latency, measured with a batch size of 1. \tool's \textit{accurate} path matches or exceeds the accuracy of the original backbone in both cases, while also achieving lower latency. This is attributed to the bottleneck injection and path adapter design, which allows the tail to begin inference at a deeper point in the network, reducing the total computation performed on the server. The \tool shared head adds only 5~ms of on-device latency, which is substantially lower than the 25~ms required by the MobileNetV2 fallback model, demonstrating that the onloaded head induces minimal on-device overhead.
\begin{table}[t]
    \caption{\small End-to-end Model Performance}
    \centering
    \setlength{\tabcolsep}{3pt}
    \begin{tabular}{llcc}
        \toprule
        \textbf{Model} & \textbf{Loc.} & \textbf{Accuracy} & \textbf{Latency} \\
        \midrule
        \tool Shared Head    & Device & ---     & \phantom{1}5ms  \\
        MobileNetV2          & Device & 73.13\% & 25ms \\
        \midrule
        ResNet152            & Server & 82.64\% & 15ms \\
        \tool ResNet152      & Server & 83.98\% & 13ms \\
        \midrule
        ConvNeXt Small       & Server & 89.35\% & 13ms \\
        \tool ConvNeXt Small & Server & 89.39\% & 11ms \\
        \bottomrule
    \end{tabular}
    \label{tab:models}
\vspace{-0.4cm}
\end{table}
\noindent
\subsection{Full System Evaluation}

\begin{figure}[t]
    \centerline{\includegraphics[scale=0.3,trim=0cm 0cm 0cm 0cm]{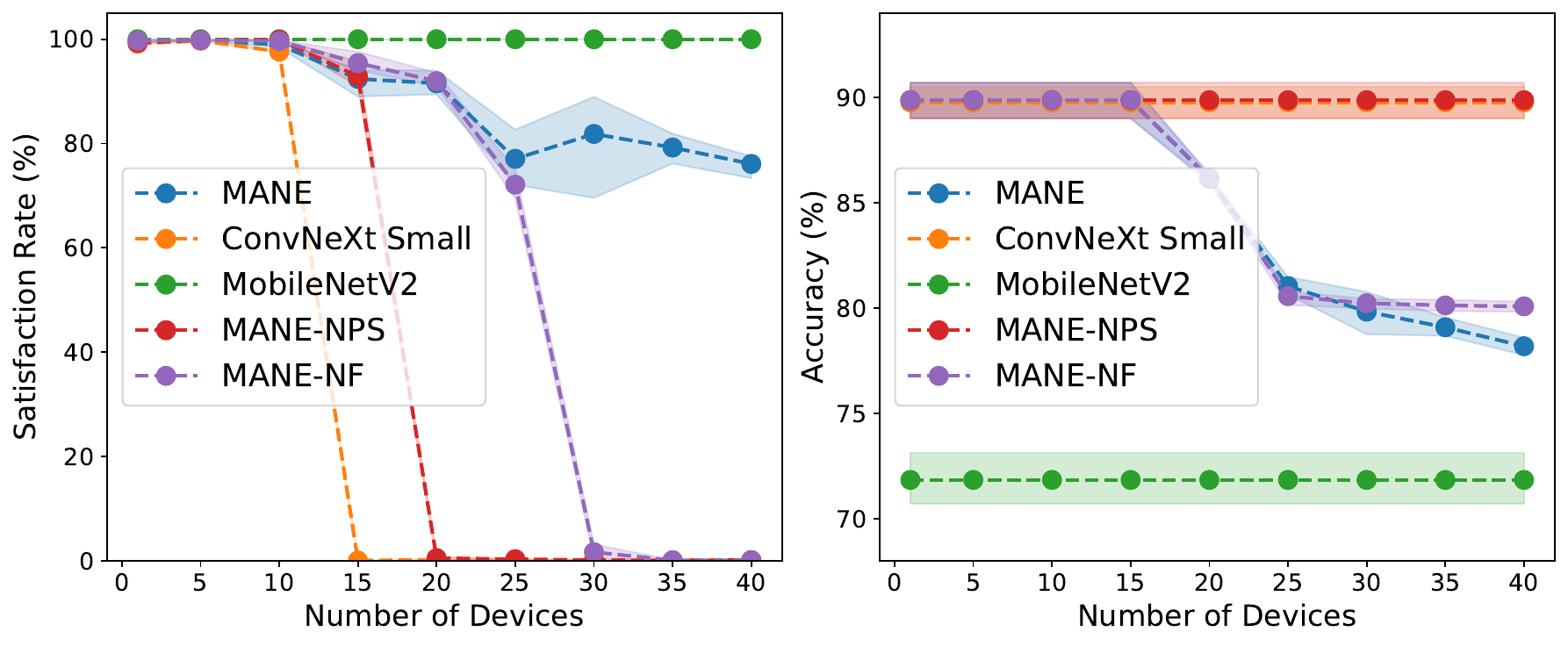}}
    \caption{\small SLO Satisfaction Rate and Accuracy for ConvNeXt Small.}
    \label{plot:convnext}
    \vspace{-0.4cm}
\end{figure}

\noindent
\textbf{SLO Satisfaction Rate \& Accuracy.}
Fig.~\ref{plot:convnext} evaluates \tool against our baselines on SLO satisfaction rate and accuracy under a challenging latency target of 100~ms. With the exception of device-only execution, all baselines eventually fail as the number of devices increases. Server-only execution using the pretrained ConvNeXt Small model collapses to a satisfaction rate of 0\% at 15 devices, with \tool-NPS and \tool-NF following at 20 and 30 devices, respectively. The difference in failure point correlates directly with the throughput of each pipeline: the original ConvNeXt Small has the lowest throughput, followed by the \textit{accurate} path used in \tool-NPS, and finally the \textit{fast} path which \tool-NF eventually settles on before failing. In contrast, \tool maintains a satisfaction rate of around 80\% even when all other baselines have completely failed, showcasing its adaptability under demanding conditions. On-device execution achieves 100\% satisfaction rate throughout, as each sample is processed in approximately 25~ms, but at a substantially lower accuracy.

On accuracy, \tool and \tool-NF begin trading off accuracy to sustain the satisfaction rate at around 20 devices. At 30 devices, \tool continues to trade accuracy via the fallback policy, while \tool-NF has exhausted its ability to do so, having already switched to the fastest path. Even under the challenging 100~ms target with 40 devices, \tool achieves a mean accuracy of 78\% with low inter-device deviation, a direct consequence of the democratic fallback rotation policy.

\begin{figure}[t]
    \vspace{0.1cm}
    \centerline{\includegraphics[scale=0.3,trim=0cm 0cm 0cm 0cm]{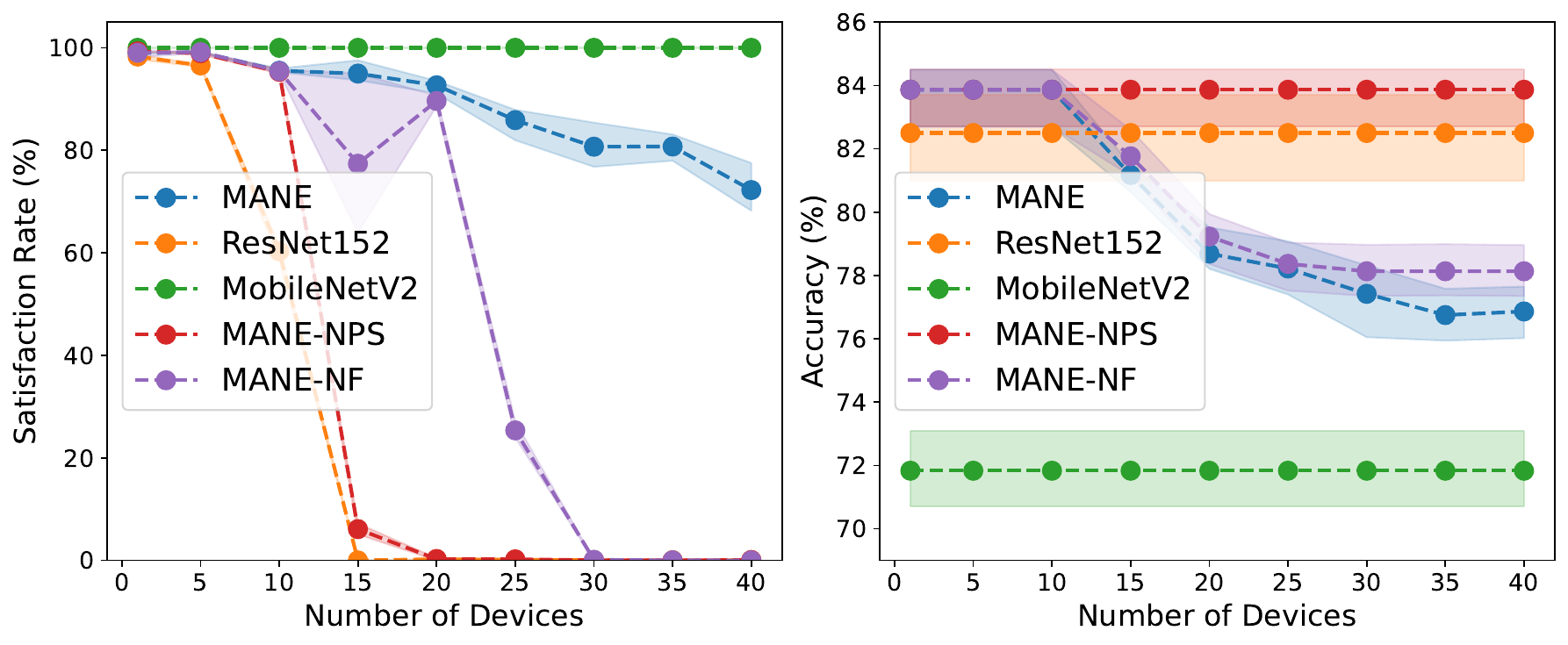}}
    \caption{\small SLO Satisfaction Rate and Accuracy for ResNet152.}
    \label{plot:resnet}
    \vspace{-0.4cm}
\end{figure}

Similar behaviour is observed in Fig.~\ref{plot:resnet}, where \tool uses ResNet152 as the server-side model. The baselines fail at 15, 20, and 35 devices, respectively, consistent with the ConvNeXt results. A more pronounced satisfaction rate drop is visible at intermediate device counts due to ResNet152's lower throughput compared to ConvNeXt Small. Despite this, \tool maintains a satisfaction rate above 75\% across all tested device counts up to 40. On accuracy, the same trade-off pattern is observed, with \tool trading more aggressively once the \textit{fast} path is insufficient and fallback is engaged. Even in the most demanding scenario, \tool achieves an accuracy of 76\%, substantially higher than on-device execution.

\begin{figure}[t]
    \centerline{\includegraphics[scale=0.3,trim=0cm 0cm 0cm 0cm]{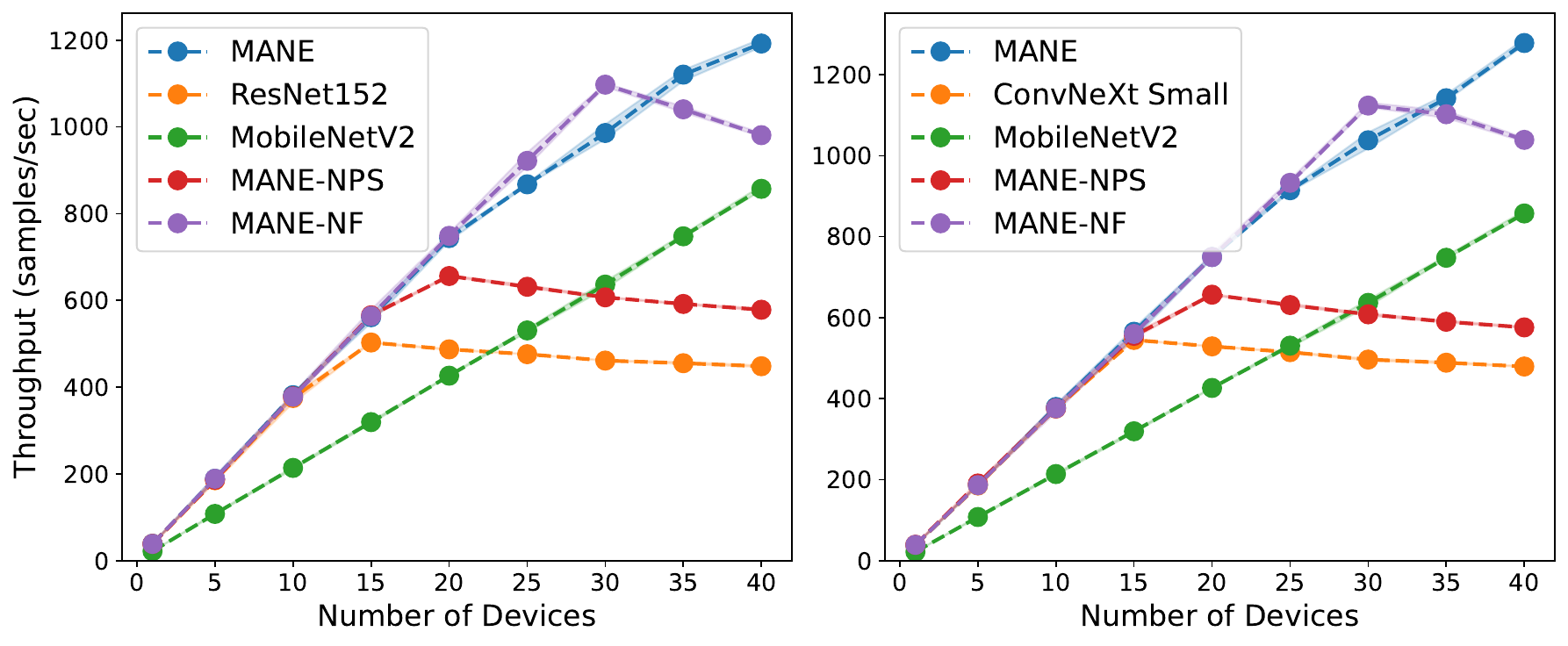}}
    \caption{\small Throughput for ResNet152 and ConvNeXt Small.}
    \label{plot:throughput}
    \vspace{-0.4cm}
\end{figure}

\noindent
\textbf{System Throughput.}
Fig.~\ref{plot:throughput} presents system throughput as the number of devices grows. Since each device contributes 1000 samples to the system, the ideal behaviour is a linear increase in throughput with the number of devices. Full server onloading and \tool-NPS plateau at around 20 devices, indicating the system has reached its maximum processing capacity. \tool-NF sustains linear throughput growth up to 30 devices before plateauing, demonstrating the benefit of the multi-path architecture in extending the system's effective capacity. \tool is the only server-assisted method that continues to increase throughput beyond this point without plateauing, reflecting the combined effect of path switching and the fallback policy in avoiding server overload.

The comparison with on-device execution is particularly instructive. Despite displaying linearly increasing throughput, on-device execution achieves substantially lower throughput values than server-assisted schemes at low device counts, as the server can process samples more efficiently by leveraging batched inference. This explains why even the more conservative baselines, \textit{i.e.}~server-only execution and \tool-NPS, outperform on-device execution in throughput when the device count is small. As device count grows enough that \tool begins reverting a significant portion to fallback, its throughput growth rate converges towards that of on-device execution, with the two curves becoming approximately parallel.
\noindent

\begin{figure}[t]
    \centerline{\includegraphics[scale=0.3,trim=0cm 0cm 0cm 0cm]{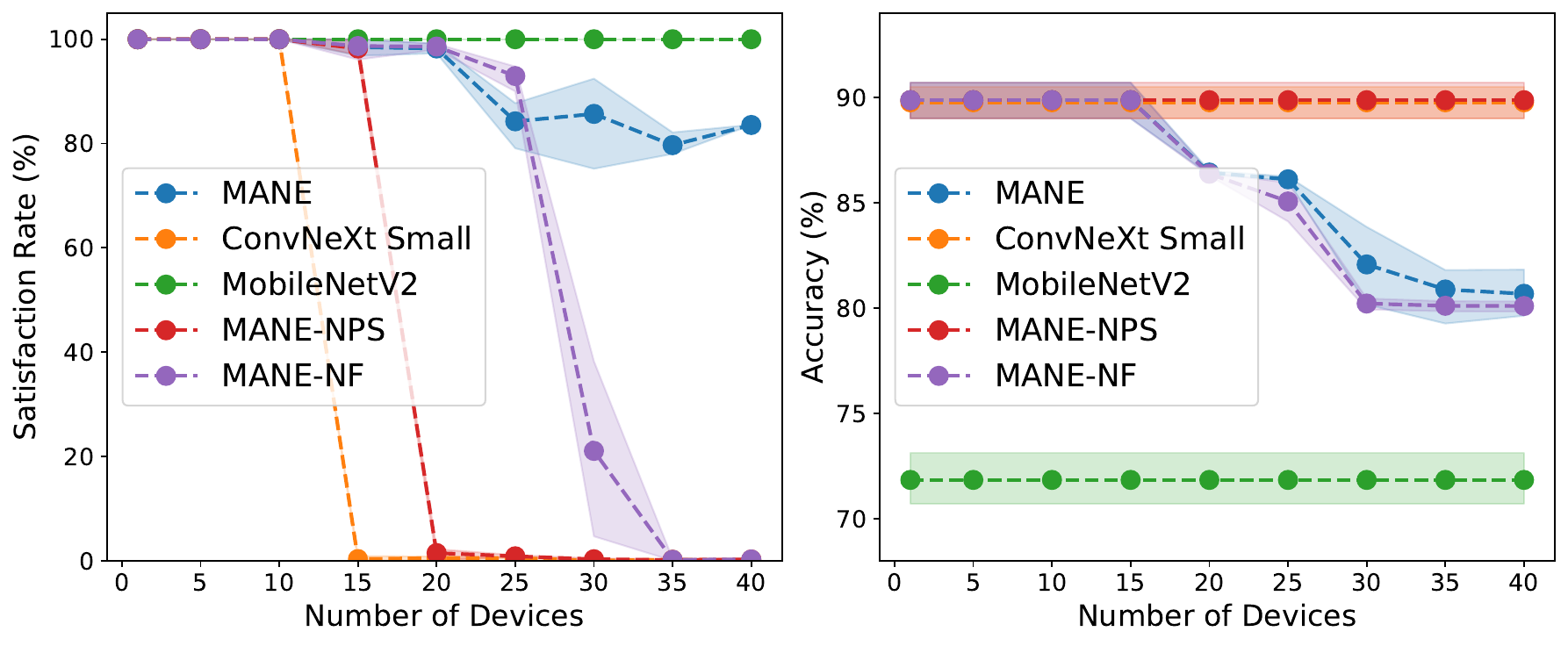}}
    \caption{\small SLO Satisfaction Rate and Accuracy with 200ms time target.}
    \label{plot:convnext200}
    \vspace{-0.4cm}
\end{figure}

\textbf{Relaxed Latency Target.}
Fig.~\ref{plot:convnext200} shows SLO satisfaction rates and accuracy for all methods under a relaxed latency target, where the scheduler can sustain a higher capacity per path and therefore trade off accuracy less aggressively. On the satisfaction rate side, server-only execution and \tool-NPS fail completely at 15 and 20 devices, respectively, consistent with the 100~ms experiment. \tool begins trading off accuracy at 25 devices and maintains satisfaction rate above 80\% throughout, mirroring its behaviour under the stricter SLO. A subtle but notable difference is observed in \tool-NF, which retains satisfaction rate of approximately 20\% at 30 devices rather than collapsing to 0\% as in the 100~ms setting, reflecting the additional headroom due to the relaxed latency budget. 

The most significant difference between the two setups is visible on accuracy. \tool leverages the relaxed latency target to preserve accuracy above 80\% across a wider range of device counts, compared to the 78\% floor observed under the 100~ms target. In this setting, \tool also consistently achieves higher accuracy than \tool-NF across all device counts.
%
%

\section{Conclusion}
\label{conclusion}
This paper presents \tool, a novel distributed inference framework that enables split-computing systems to operate effectively in demanding multi-device edge environments. By equipping the server with a multi-path tail architecture and a Joint Head Network Distillation scheme, we introduce a dynamic accuracy--throughput trade-off that existing static onloading approaches cannot provide. \tool continuously adapts the active inference path in response to server load, while its equitable fallback policy ensures that no single device monopolises server resources, sustaining high satisfaction rates and accuracy across a growing number of concurrent devices. \tool's adaptability and scalability represent a step towards making split computing a viable and robust solution for the intelligent indoor environments of the future.

\bibliographystyle{IEEEtran}
{\footnotesize
\bibliography{references}
}

\end{document}